\documentclass[onecolumn, 12pt, journal, letterpaper]{IEEEtran}
\usepackage{multirow}
\usepackage{amsmath, amssymb, setspace, cite}
\usepackage{graphicx}
\usepackage{epsfig}
\usepackage{verbatim}
\usepackage{latexsym}
\usepackage{rotating}
\usepackage{array}
\usepackage{fixltx2e}
\usepackage{url}
\begin{document}
\title{Relay Selection and Power Allocation in Cooperative Cellular Networks}
\author{\IEEEauthorblockN{Sachin Kadloor and Raviraj Adve} \\
\IEEEauthorblockA{Department of Electrical and Computer Engineering, University of Toronto\\
10 King's College Road, Toronto, ON M5S 3G4, Canada \\
Email: \texttt{\{skadloor, rsadve\}@comm.utoronto.ca}}}

\maketitle

\begin{abstract}
%\boldmath
We consider a system with a single base station communicating with
multiple users over orthogonal channels while being assisted by
multiple relays. Several recent works have suggested that, in such a
scenario, selection, i.e., a single relay helping the source, is the
best relaying option in terms of the resulting complexity and overhead.
However, in a multiuser setting, optimal relay assignment is a
combinatorial problem. In this paper, we formulate a related convex
optimization problem that provides an extremely tight upper bound on
performance and show that selection is, almost always, inherent in the
solution. We also provide a heuristic to find a close-to-optimal relay
assignment and power allocation across users supported by a single
relay. Simulation results using realistic channel models demonstrate
the efficacy of the proposed schemes, but also raise the question as to
whether the gains from relaying are worth the additional costs.
\end{abstract}

\doublespacing

\section{Introduction}

In distributed wireless systems wherein each node possesses only a
single antenna, relays can be used to provide spatial diversity and
combat the impact of fading. Relaying has been an extremely active
research area, especially since Sendonaris et al.,
in~\cite{SendonarisErkip}, proposed the idea of user cooperation
wherein mobile users cooperate by relaying each others' data. Many
cooperation schemes have now been studied, e.g.,~\cite{SendonarisErkip,
LanemanTseWornell, Laneman_dstc, HunterNosratinia_coded}. The work
in~\cite{LanemanTseWornell} and~\cite{Laneman_dstc} proposed
repetition-based cooperation schemes including fixed
amplify-and-forward (AF) and decode-and-forward (DF) using orthogonal
channels (time/frequency slots). In networks with multiple relays, the
traditional strategy has been to let all the relays forward their
messages to the destination. However, having relays transmit on
orthogonal bands is bandwidth inefficient. A proposed alternative is to
use distributed space-time codes (DSTC)~\cite{Laneman_dstc}; however,
this requires symbol level synchronization, which is difficult to
implement over a distributed network. It has recently been shown that
most of the benefits of cooperative diversity can be achieved with
minimum overhead if a single `best' relay cooperates with the source.
This scheme is referred to as selection
cooperation~\cite{BletsasKhistiReedLippman,Ela_jnl} and has now been
investigated in various
contexts~\cite{BletsasKhistiReedLippman,Ela_jnl,
ZhaoAdveLim_SelectionAFPowerAlloc, MichalopoulosKaragiannidis-TWC,
JosephineICCSelectionBhattacharya}.

In the case of a single source-destination pair, choosing the best
relay is fairly straightforward and solved for both
DF~\cite{BletsasKhistiReedLippman, Ela_jnl} and
AF~\cite{ZhaoAdveLim_SelectionAFPowerAlloc} relaying. In both cases,
the best relay is the one that contributes the most to the output
signal-to-noise ratio (SNR). The selection gets significantly more
complicated in the more practical case of multiple information
flows~\cite{Ela_jnl}. Because a relay must now divide its available
power between all flows it supports, a relay that is best for a single
flow may not remain the best overall and relay selection becomes a
combinatorial problem. In~\cite{Ela_jnl}, the authors present ad hoc
approaches to approximate the optimal solution with limited complexity,
without addressing resource allocation.

In relay networks, an independent research theme is that of resource
allocation, including power allocation,
e.g.,~\cite{LiangVeeravalliPoor-MaxMin, YenerPowerAlloc} amongst many.
Optimal allocation makes best use of the limited available power
resources. However, to our knowledge there has not been any work
jointly considering relay assignment and power allocation in the
context of multiple source, multi-relay, especially cellular, networks.

Our system model comprises a single base station communicating to
multiple users being assisted by a few dedicated relays. The users are
to be assigned to the relays. The relays have limited power which must
be divided among the users they support. Relaying in the context of a
cellular wireless network has received limited
attention~\cite{ReznikHSDPA}. In Section~\ref{probform}, we develop an
optimization problem for optimal relay assignment and power allocation
at the relays. We try to answer the question, \textit{what relay
assignment and power allocation scheme maximizes the sum rate and what
scheme maximizes the minimum rate to all the users}? Obtaining
solutions to these requires exponential complexity.

The main theoretical contribution of this paper is in
Section~\ref{probform}, where we derive upper bounds to the rates and
show that these bounds form a convex optimization problem for both
figures of merit. We use the resulting Karush-Kuhn-Tucker (KKT)
conditions to illustrate why the bound is tight and then derive a
simplified, tight, lower bound. In Section~\ref{res}, we simulate a
cellular network, using the COST-231 model, to study the performance
gains in a relay assisted network over a traditional single base
station system. Interestingly, while the gains are significant, the
results leave open the question of whether these gains adequately
compensate for the additional infrastructure costs of a relay-assisted
cellular system.

In terms of the available literature, our formulation is similar to
tone assignment in orthogonal frequency division multiple access
(OFDMA) systems. This is the problem of assigning users to tones to
maximize a certain metric, subject to the constraint that no tone is
assigned to two different users. In~\cite{ofdmbitpoweralloc}, Wong et
al.~solve the problem of minimizing the total transmission power in a
multi-user OFDM system. The frequency spectrum is divided into discrete
frequency bins and a convex relaxation technique is used to solve the
discrete bin-assignment problem. Rhee and Cioffi~\cite{ofdm_maxmin}
solve the problem of assigning users to tones on the downlink of a
OFDMA based communication system with the objective of maximizing the
minimum rate to users.

While our approach is similar to that taken in these papers; the roles
of `user' and `tone' in these papers is played, respectively, by
`relay' and `user' here. However, there are some significant
differences, most importantly in the power allocation step. There a
single power allocation is required across all tones, here each relay
must meet a power constraint Furthermore, our problem formulation
allows us to analyze the conditions under which the bounds are tight
unlike the other works, wherein the authors state that the bound gets
tight as the number of tones approaches infinity, but prove this for
only the two user case.

This paper is organized as follows. In Section~\ref{sysdesc}, we
describe the system model in some detail. In Section~\ref{probform}, we
then formulate the optimization problem and the upper bound to each of
the two rates and illustrate why the bounds are tight. In
Section~\ref{res}, we illustrate this through simulations and use the
bound to analyze the performance of relay assisted cellular networks.
The paper wraps up with some conclusions in Section~\ref{sec:conc}.

\section{System Model}\label{sysdesc}

\begin{figure}
\centering
 \includegraphics[scale =0.6]{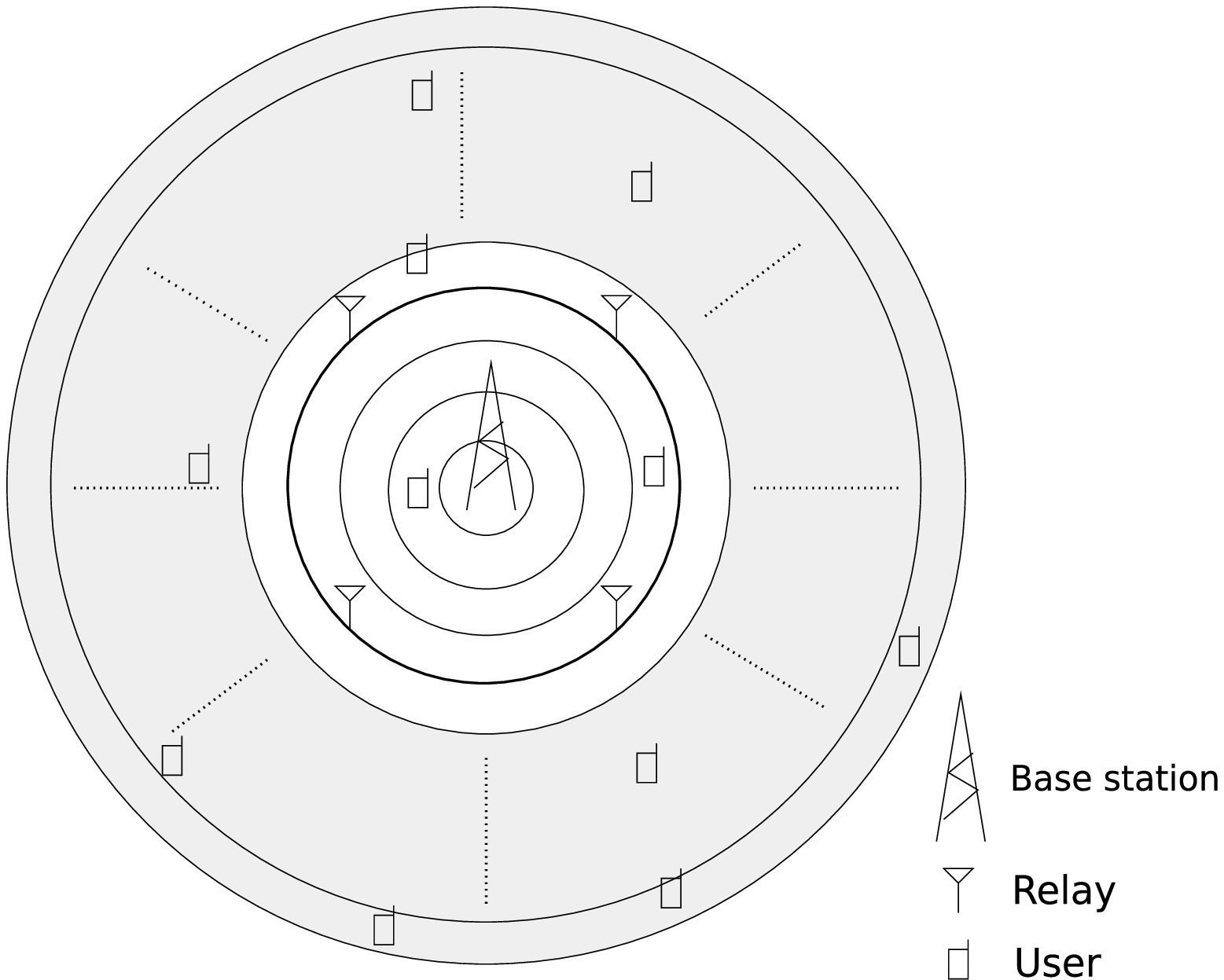}
\caption{A relay aided cellular network}
\label{sysmodel}
\end{figure}

Our system model consists of a cellular network with a single BS,
communicating with $K$ users, and assisted by $J$ relays, as shown in
Figure~\ref{sysmodel}. Each of the users is assigned an orthogonal
channel, over which the BS-to-user and the relay-to-user communications
take place. The users are frequency division multiplexed, although the
results here also apply to the case of time division multiplexing. The
relays in the system are fixed wireless terminals, installed solely to
aid the BS-user communication. The relays use the DF protocol with the
same codebook as the transmitter.

The communication between the BS and a user happens over two time
slots. In the first time slot the BS transmits, while the relays and
the user try to decode the message. In the second time slot, one of the
relays, chosen \emph{a priori}, re-encodes and then transmits the
information it has decoded in the first time slot. The user uses the
messages received in the two time slots to decode the transmitted
information.

Suppose that user $k$ (denoted as $d_k$) is allotted to relay-$j$
($r_j$). For a system as described above, the maximum rate at which the
BS can communicate with the receiver with the help of the relay
is~\cite{Ela_jnl}
\begin{eqnarray}
I_{d_k} & = & \min\left(I_{sr_j},I_{sr_jd_k}\right), \label{eq:MinOfTwo} \\
I_{sr_j} & = & \frac{1}{2}\log_2\left( 1+ \texttt{SNR}_s|h_{sr_j}|^2\right), \label{eq:RateISR}\\
I_{sr_jd_k} & = &  \frac{1}{2}\log_2\left( 1+ \texttt{SNR}_s|h_{sd_k}|^2+
                            \texttt{SNR}_{r}\alpha_{jk}|h_{r_jd_k}|^2\right), \label{eq:RateISRD}
\end{eqnarray}
where, $\texttt{SNR}_s$ and $\texttt{SNR}_r$ are, respectively, the
ratios of the transmit power at the BS (denoted as $s$) and relay to
the noise power at the receiver. $h_{sr_j}$ is the channel between the
BS and relay $j$, denoted by $r_j$, similarly $h_{r_jd_k}$ is the
channel between relay $r_j$ and destination $d_k$. Finally,
$\alpha_{jk}$ is the fraction of the total relay power used to
communicate with user $k$. The factor of 1/2 accounts for the fact that
the BS-user communication happens over two time slots. $I_{sr_j}$ is
the rate at which the source can communicate with relay-$j$ while
$I_{sr_jd_k}$ is the maximum rate at which the source can communicate
to user $k$ with the help of the relay. Equation~\eqref{eq:MinOfTwo}
ensures that both the relay and the user can decode the message.

The channels between the BS, relays and users are modeled using the
COST-231 model as recommended by the IEEE 802.16j working
group~\cite{80216jrelay}. The model includes the path loss, large-scale
fading (a log-normal variable) and small-scale fading modeled as Rician
random variable for line-of-sight (LoS) communication and Rayleigh
random variable for non-LoS communication. When the BS and relays are
both placed at some height above the ground, the fading has a LoS
component. The existence of this component is crucial since it suggests
that all relays will be able to decode a source codeword; hence the
factor limiting the overall rate is the second term of
Eqn.~\eqref{eq:MinOfTwo}, $I_{sr_jd_k}$, significantly simplifying the
problem at hand.

\section{Problem formulation and solution}\label{probform}
As described in the previous section, every user is assigned one of the
$J$ relays. This paper deals with optimizing this assignment to
maximize two metrics of interest, the sum rate to all the users and the
minimum of all the rates. In maximizing the sum rate (equivalently the
average rate), the objective function is
\begin{equation}\label{minsum}
\sum_{k=1}^{K} I_{d_k} = \sum_{k=1}^{K}  \min\left( I_{sr(d_k)},I_{sr(d_k)d_k} \right),
\end{equation}
while in maximizing the minimum rate, the objective function is given
by
\begin{equation}\label{min}
\min_k \left\{ I_{d_k} \right\} =
        \min_k \left\{ \min\left( I_{sr(d_k)},I_{sr(d_k)d_k} \right) \right\},
            \hspace*{0.2in} k = 1, \dots, K,
\end{equation}
where, in both cases, $r(d_k)$ is the relay assigned to user $k$.

In practice, the number of users, $K$, will be much larger than the
number of relays, $J$. Hence, a single relay will likely be required to
support multiple users, and to meet its power constraint it must divide
its power amongst these users. Thus, our objective is now two fold,
one, finding the relay assignment scheme, and two, once the assignment
is done, distributing powers at each of the relays amongst the users it
supports.

To formulate a tractable problem, in this paper we investigate
simplified versions of the above problems. As mentioned earlier, in a
cellular network, the data rate bottleneck is the compound
source-relay-destination channel, the second term in
Eqn.~\eqref{eq:MinOfTwo}. We assume that
\begin{equation}
I_{sr_j} > I_{sr_jd_k} \quad \forall j,k,
\label{assump}
\end{equation}
and hence $ \min\left( I_{sr(d_k)},I_{sr(d_k)d_k} \right) =
I_{sr(d_k)d_k}. $

In Section~\ref{res}, we justify the validity of this assumption. Note
that in spite of the assumption, the solution is not immediate. The
fact that the relays divide their power amongst the users they support,
makes the relay assignment an integer programming problem with the
attendant exponential complexity.

\subsection{Max sum rate}

The sum rate measures the maximum throughput delivered by the base
station. For the sake of brevity, let $c_k$ represent
$\texttt{SNR}_s|h_{s d_k}|^2$ and $p_{jk}$ represent
$\texttt{SNR}_r|h_{r_{j} d_k}|^2, \; j=1,2,\ldots, J$. Let
$\alpha_{jk}$ be the fraction of the power of relay-$j$, used to
communicate to user $k$. The optimization problem maximizes the sum
rate to all the users subject to two constraints: only a single relay
helps each user and each relay must meet a power constraint. The formal
optimization problem is, therefore,
\begin{eqnarray}
\max_{\left\{\alpha_{jk}\right\}} R & = & \hspace*{-0.18in} \max_{\left\{\alpha_{jk}\right\}}
                            \sum_{k=1}^{K} \frac{1}{2} \log_2 \left( 1+c_k +
                                   \sum_{j=1}^J p_{jk} \alpha_{jk} \right), \label{mainobj} \\[6pt]
\mathrm{such\ that} & \forall k, & \alpha_{jk}\alpha_{lk} = 0, \hspace*{0.3in}
                          j\neq l,j,l \in \left\{1, 2, \dots, J\right\}, \label{const} \\[4pt]
& & \sum_{k=1}^K \alpha_{jk} = 1 \hspace*{0.3in} \forall \hspace*{0.1in} j, \label{pow} \\[4pt]
& & \alpha_{jk} \geq 0, \label{positivepower}
\end{eqnarray}
where the objective function assumes the relay uses the same codebook
as the source. Equation~\eqref{const} enforces the selection rule
allowing only one $\alpha_{jk}$ term to be non-zero for all relays. The
remaining two constraints force the power allocated to be positive and
to meet a power constraint. The constraint in Eqn.~\eqref{pow} can also
be written as an inequality constraint, $\sum_{k=1}^K \alpha_{jk} \leq
1, \forall j$. The solution to the optimization problem in either case
is the same because the objective function is an increasing function of
the powers, $\alpha_{jk}$s. We cannot use the usual gradient based
methods to maximize the objective function in Eqn.~\eqref{mainobj}.
Note that an inherent assumption is that the BS has knowledge of the
parameters that define the problem. How this information is conveyed to
the BS is beyond the scope of this paper.

The solution to the optimization problem in
Eqns.~\eqref{mainobj}-\eqref{positivepower} is complicated by the
constraint in Eqn.~\eqref{const}. An exhaustive search to find the
solution would involve the following: for a given relay assignment,
solving $J$ water-filling problems corresponding to the power
allocation at each of the relays. We need do this for every relay
assignment and find the maximum of them. Each of the users can be
assigned to any of the relays, hence, all  $J^K$ possible relay
assignments must be tested. Doing so is impossible for realistic values
of $J$ and $K$. We therefore explore tractable approximate
formulations.

The objective function of the optimization problem in
Eqns.~\eqref{mainobj}-\eqref{positivepower} is concave and the
constraints, other than the one in  \eqref{const}, are affine. Our
strategy to solve the optimization problem in hand is to ignore the
constraints in Eqn.~\eqref{const} and maximize the objective function
subject to the power constraints alone:
\begin{eqnarray}
& & \hspace*{-0.4in} \min_{\left\{\alpha_{jk}\right\}}
              -\sum_{k=1}^{K} \frac{1}{2} \log_2 \left( 1+c_k +
                     \sum_{j=1}^J p_{jk} \alpha_{jk} \right), \label{mainobj1}\\[6pt]
\mathrm{such\ that}
& & \sum_{k=1}^K \alpha_{jk} - 1 = 0 \hspace*{0.3in} \forall \hspace*{0.1in} j,
\label{pow1}\\[6pt]
& & -\alpha_{jk} \leq 0\label{positivepower1}.
\end{eqnarray}
Since we ignore a constraint, the solution so obtained will be an upper
bound to the maximum sum rate achieved by selection. Note that since
this minimization problem is now convex, solving this simplified
problem is fairly straightforward, e.g., using interior point
methods~\cite{Boydbook}. The computational complexity involved in
solving the optimization problem is polynomial in $K$ and $J$, and the
problem is, hence, tractable for practical values of $K$ and $J$.  We
use \texttt{CVX}, a package for specifying and solving convex programs
\cite{CVX,DCP}.

We now proceed to show that although we did not impose the selection
rule explicitly, the solution to the optimization problem has the
property that, for most $k$, $\alpha_{jk}\alpha_{lk} = 0, j\neq l,j,l
\in \left\{1, 2, \dots, J\right\}$ . This means, when the power is
optimally allocated, most users receive power from only \emph{one} of
the relays.

\noindent \emph{Tightness of Bound}: The objective and the constraint
functions are differentiable and the constraint conditions satisfy
Slater's condition\cite{Boydbook}. To show this, consider one possible
choice for the power vectors, $\alpha_{jk}=1/K$. This meets the sum
power constraint with equality and the constraint on positive power
with strict inequality. For a convex optimization problem with
differentiable objective and constraint functions, which also satisfy
Slater's condition, the solution to the optimization problem satisfies
the KKT conditions~\cite{Boydbook}.

Let us characterize the set of solutions to the optimization problem.
For the sake of clarity, we start with the case with two relays. In
such a case, the Lagrangian of the minimization problem is given by
\begin{eqnarray}
\mathcal{L}(\left\{\alpha_{1k},\alpha_{2k}\right\};
    \{\lambda^{1}_k\},\{\lambda^{2}_k\}, \nu_1,\nu_2) & = &
                -R - \sum_{k=1}^{K}\lambda^{1}_k \alpha_{1k}
                - \sum_{k=1}^{K}{\lambda^{2}_k} \alpha_{2k} \nonumber\\
        & & + \nu_1 \left( \sum_{k=1}^{K} \alpha_{1k} -1\right) +
                            \nu_2 \left( \sum_{k=1}^{K} \alpha_{2k} -1\right),
\end{eqnarray}
where $\lambda^{1}_k$ and $\lambda^{2}_k, k = 1,2,\dots, K$ are the
Lagrange multipliers associated with the constraint on positive power,
and $\nu_1$ and ${\nu_2}$ are the Lagrange multipliers associated with
the constraint on the total power at the two relays. Any solution to
the optimization problem satisfies the KKT conditions, which are,
\begin{eqnarray}\label{kkt}
\frac{p_{1k}}{1+c_k+ \sum_{i=1}^2 p_{ik} \alpha_{{i} k} } + \lambda^{1}_k =
            \nu_1, \quad \lambda^{1}_k \alpha_{1k} = 0, \quad \lambda^{1}_k \geq 0,\\
\frac{p_{2k}}{1+c_k+ \sum_{i=1}^2 p_{ik} \alpha_{{i} k}} + {\lambda^{2}_k} =
            {\nu_2 }, \quad {\lambda^{2}_k} \alpha_{2k} = 0, \quad {\lambda^{2}_k}\geq 0.
\end{eqnarray}

Now suppose for some $i \in \{1,2,\ldots,K\}$, $\alpha_{1i}$ and
$\alpha_{2i}$ are both non-zero, then the conditions $\lambda^{1}_i
\alpha_{1i} = 0 $ and $\lambda^{2}_i \alpha_{2i} = 0$ dictate that
$\lambda^{1}_i$ and $\lambda^{2}_i$ are both zero. From the KKT
conditions it follows that
\begin{equation}\label{const1}
\frac{\nu_1}{p_{1i}}=\frac{{\nu_2}}{p_{2i}}.
\end{equation}
Similarly if $\alpha_{1j}$ and $\alpha_{2j}$ are both non-zero for some
$j\in \{1,2,\ldots,K\}$, then
\begin{equation}\label{const2}
\frac{\nu_1}{p_{1j}}=\frac{{\nu_2}}{p_{2j}}.
\end{equation}
Unless $p_{1i}/p_{2i}=p_{1j}/p_{2j}$, Eqns.~\eqref{const1}
and~\eqref{const2} cannot simultaneously be true. In the current
problem, $p_{jk}$ represent the powers of the channels between the
relays and the users. If they are independent continuous random
variables, as is the case with the wireless channels, then the
probability that $p_{1i}/p_{2i}=p_{1j}/p_{2j}$ is zero. Hence,
\emph{when the power is optimally allocated, \emph{at most} one of the
$K$ ($\alpha_{1k},\alpha_{2k}$) pairs has two non-zero entries} and
$K-1$ of the  pairs have at most one non-zero entry. This indicates
that the selection rule, $\left(\alpha_{1k}\alpha_{2k}=0, \forall
k\right)$, which we did not explicitly impose, is true for at least all
but one of the $K$ users. Hence, the solution obtained by ignoring
Eqn.~\eqref{const} comes quite close to the solution to the original
optimization problem in Eqns.~\eqref{mainobj}-\eqref{positivepower}.

For the case of three relays, the KKT conditions are:
\begin{eqnarray}
\frac{p_{1k}} {1+c_k+ \sum_{i=1}^{3} p_{ik} \alpha_{{i} k} } + \lambda^1_k =
            \nu_1, \quad \lambda^1_k \alpha_{1k} = 0, \hspace{0.04cm} \lambda^1_k \geq 0, \\
\frac{p_{2k}}{1+c_k+ \sum_{i=1}^{3} p_{ik} \alpha_{{i} k} } + \lambda^2_k =
            \nu_2, \quad \lambda^2_k \alpha_{2k} = 0, \hspace{0.04cm} \lambda^2_k \geq 0, \\
\frac{p_{3k}}{1+c_k+ \sum_{i=1}^{3} p_{ik} \alpha_{{i} k} } + \lambda^3_k =
            \nu_3, \quad \lambda^3_k \alpha_{3k} = 0, \hspace{0.04cm} \lambda^3_k \geq 0,
\end{eqnarray}
where, for $k\in\{1,2,\ldots,K\}$, $\lambda^{1}_k$, $\lambda^{2}_k$ and
$\lambda^{3}_k$ are the Lagrangian multipliers associated with the
constraint on positive power, and $\nu_1$, $\nu_2$ and $\nu_3$ are the
Lagrangian multipliers associated with the constraint on total power.
In the solution to the optimization problem, we wish to find the
maximum number of triplets ($\alpha_{1i}$, $\alpha_{2i}$,
$\alpha_{3i}$), in which more than one entry is non-zero. We do this by
analyzing different possibilities for the solution. Suppose that in the
solution, for some $i$, ($\alpha_{1i}$, $\alpha_{2i}$, $\alpha_{3i}$)
are all non-zero (user $i$ receives power from all relays), then,
\begin{equation}\label{three1}
\frac{\nu_1}{p_{1i}}=\frac{{\nu_2}}{p_{2i}}=\frac{{\nu_3}}{p_{3i}}.
\end{equation}
Now, for some $j$, if $\alpha_{1j}$ and $\alpha_{2j}$ are non-zero,
then, along with Eqn. \eqref{three1}, this would imply that
$p_{1i}/p_{2i}=p_{1j}/p_{2j}$, which occurs with probability zero.
Hence, if the solution to the optimization problem has one triplet with
all non-zero entries, then all other triplets can have only one
non-zero entry, i.e., selection is imposed on all other users.

Now suppose that in the solution, for no $i$, ($\alpha_{1i}$,
$\alpha_{2i}$, $\alpha_{3i}$) are all non-zero.  Without loss of
generality, suppose for some $j$, $\alpha_{1j}$ and $\alpha_{2j}$ are
non-zero, and for some $k$, $\alpha_{2k}$ and $\alpha_{3k}$ are
non-zero, then,
\begin{eqnarray}\label{three2}
\frac{\nu_1}{p_{1j}}=\frac{{\nu_2}}{p_{2j}}, \qquad \frac{{\nu_2}}{p_{2k}}=\frac{{\nu_3}}{p_{3k}}.
\end{eqnarray}
These two equations imply that in all other three-tuples ($\alpha_{1k},
\alpha_{2k}, \alpha_{3k}$), only one of the entries is non-zero.  This
is because, if for some $l$, $\alpha_{1l}$ and $\alpha_{3l}$ are
non-zero, then,   \eqref{three2} would imply,
$p_{1l}/p_{3l}=p_{2j}p_{1i}/p_{3j}p_{2i}$, which occurs with
probability zero.  Hence, for the case of three relays, at most two of
triplets can have more than one non-zero entry. Like with the case of
two relays, when the power is allocated optimally, the selection rule
is followed in most of the triplets.

Generalizing this to $J$ relays, when the power is allocated optimally,
at most $J-1$ of the $J$-tuples ($\alpha_{1k}, \alpha_{2k}, \ldots,
\alpha_{Jk}$) can have more than one non-zero entry. This indicates
that if $K \gg J-1$, as expected in practice, a large fraction of the
users are guaranteed to receive power from only one relay.

To summarize, we have shown that the power allocation matrix,
$[\alpha_{jk}]_{J\times K}$ is \emph{sparse}. Most of the rows of the
matrix have only one non-zero entry. If all the rows of the matrix had
at most a single non-zero entry, then we would have obtained the
solution to the optimization problem given by
Eqns.~\eqref{mainobj}-\eqref{positivepower}. A simple heuristic to find
that solution, then, is to explicitly impose selection: assign users
receiving power from multiple relays to the relays that allot the
maximum power.
\begin{eqnarray}
r(d_k) = r_m &  \mathrm{if}  &  m = \arg \max_j \left\{\alpha_{jk}p_{jk} \right\}
\end{eqnarray}
If there are multiple relays which allot the same maximum power, assign
the user to any one of them arbitrarily. Once this relay assignment is
done, $J$ water-filling problems can be solved for the power
distribution at each of the relays. However, we can also re-use the
power allocation vector derived from the earlier step. Construct
the matrix $[\alpha^{'}_{jk}]_{J\times K}$ as follows: for each
$k\in\{1,2,\ldots K\}$,
\begin{eqnarray}
\alpha^{'}_{mk}=\alpha_{mk} \quad \alpha^{'}_{jk}= 0& \forall j \ne m.
\end{eqnarray}
The matrix of the power allocation vectors $[\alpha^{'}_{jk}]_{J\times
K}$ meet the constraints given by Eqns.~\eqref{const} and
\eqref{positivepower} and satisfy $\sum_{k=1}^{K}\alpha^{'}_{jk} \leq
1, \forall j$. It is hence a lower bound to the solution to the
optimization problem given by
Eqns.~\eqref{mainobj}-\eqref{positivepower}. We avoid a second round of
optimization because, as we shall see, the upper and lower bounds are
already indistinguishable.

\subsection{Max sum rate with a minimum rate constraint }
Maximizing the sum rate does not ensure any fairness with respect to
the distribution of power. In a cellular network, a more practical
metric might be maximizing the sum rate while guaranteeing a minimum
rate to each user. Formally, the optimization problem is,
\begin{eqnarray}
\max_{\left\{\alpha_{jk}\right\}} R & \hspace*{-0.18in}= &
                \hspace*{-0.18in} \min_{\left\{\alpha_{jk}\right\}}
                            - \sum_{k=1}^{K} \frac{1}{2} \log_2 \left( 1+c_k +
                     \sum_{j=1}^J p_{jk} \alpha_{jk} \right), \label{maxwithminobj} \\[4pt]
\mathrm{such\ that} & \forall k, & \alpha_{jk}\alpha_{lk} = 0, \hspace*{0.03in}
                j\neq l,j,l \in \left\{1, 2, \dots, J\right\} \label{maxwithminconst} \\[4pt]
& & \sum_{k=1}^K \alpha_{jk} - 1 =0 \hspace*{0.3in} \forall \hspace*{0.1in} j,
                                                                \label{maxwithminpow} \\[4pt]
& & -\alpha_{jk} \leq 0, \label{maxwithminpositivepower}\\[10pt]
& & \hspace*{-.28in}R_k - \frac{1}{2} \log_2 \left( 1 + c_k +
                    \sum_{j=1}^J p_{jk} \alpha_{jk} \right) \leq 0, \label{maxwithminrate}
\end{eqnarray}
where $R_k$ is the rate guaranteed to user $k$. Suppose we ignore the
constraint given in Eqn.~\eqref{maxwithminconst}, the Lagrangian of the
resulting optimization problem, for the case of $J=2$ relays is given
by :
\begin{eqnarray}
\mathcal{L}(\left\{\alpha_{1k},\alpha_{2k}\right\};
    \{\lambda^{1}_k\},\{\lambda^{2}_k\}, \nu_1,\nu_2,\{\gamma_k\}) & = &
                -R - \sum_{k=1}^{K}\lambda^{1}_k \alpha_{1k}
                - \sum_{k=1}^{K}{\lambda^{2}_k} \alpha_{2k}  \nonumber\\
& & + \nu_1 \left( \sum_{k=1}^{K} \alpha_{1k} -1\right) +
\nu_2 \left( \sum_{k=1}^{K} \alpha_{2k} -1\right)\nonumber\\
& &  -\sum_{k=1}^{K}{\gamma_k} \left( R_k -
        \frac{1}{2} \log_2 \left( 1 + c_k + \sum_{j=1}^2 p_{jk} \alpha_{jk} \right)\right),
\end{eqnarray}
where $\lambda^{1}_k$ and $\lambda^{2}_k, k = 1,2,\dots, K$ are the
Lagrange multipliers associated with the constraint on positive power.
 $\nu_1$ and ${\nu_2}$ are the Lagrange multipliers associated with
the constraint on the total power at the two relays, and $\gamma_k, \; k =
1,2,\dots, K$ are the Lagrange multipliers associated with the
constraint on the minimum rate. The solution, if it exists, satisfies
the KKT conditions, which are,
\begin{eqnarray}\label{kktmaxwithmin}
\frac{p_{1k}(2-\gamma_k)}{2\left(1+c_k+ \sum_{i=1}^2 p_{ik} \alpha_{{i} k} \right)}
                                                    + \lambda^{1}_k = \nu_1,
\quad \lambda^{1}_k \alpha_{1k} = 0, \quad \lambda^{1}_k \geq 0,  \\[10pt]
\frac{p_{2k}(2-\gamma_k)}{2\left(1+c_k+ \sum_{i=1}^2 p_{ik} \alpha_{{i} k} \right)}
                                                    + \lambda^{2}_k = \nu_2,
\quad {\lambda^{2}_k} \alpha_{2k} = 0, \quad {\lambda^{2}_k}\geq 0,\\[10pt]
\hspace*{-.1in}{\gamma_k} \left( R_k - \frac{1}{2} \log_2 \left( 1 + c_k +
        \sum_{j=1}^2 p_{jk} \alpha_{jk} \right)\right) =0,\quad \gamma_k \geq 0.
\end{eqnarray}
Suppose for some $i \in \{1,2,\ldots,K\}$, $\alpha_{1i}$ and
$\alpha_{2i}$ are both non-zero, then the conditions $\lambda^{1}_i
\alpha_{1i} = 0 $ and $\lambda^{2}_i \alpha_{2i} = 0$ dictate that
$\lambda^{1}_i$ and $\lambda^{2}_i$ are both zero, and from the KKT
conditions it follows that
\begin{equation}\label{constmaxwithmin1}
\frac{\nu_1}{p_{1i}}=\frac{{\nu_2}}{p_{2i}}.
\end{equation}
Similarly if $\alpha_{1j}$ and $\alpha_{2j}$ are both non-zero for some
$j\in \{1,2,\ldots,K\}$, then
\begin{equation}\label{constmaxwithmin2}
\frac{\nu_1}{p_{1j}}=\frac{{\nu_2}}{p_{2j}},
\end{equation}
hence, like the case with the max sum rate metric, when the power is
optimally allocated, at most one of the $K$ $(\alpha_{1k},\alpha_{2k})$
pairs can have more than one non-zero entry. To generalize this result,
with $J$ relays at most $J-1$ of the $J$-tuples ($\alpha_{1k},
\alpha_{2k}, \ldots, \alpha_{Jk}$) have more than one non-zero entry.
The additional constraint on minimum rate given in
Eqn.~\eqref{maxwithminrate} does not alter this property of the
solution.

A lower bound to the solution of the optimization problem, can be
formulated like before. Explicitly impose selection, by assigning users
receiving power from multiple relays to the relays that allot the
maximum power,
\begin{eqnarray}
r(d_k) = r_m &  \mathrm{if}  &  m = \underset{j}{\mathrm{argmax}} \left\{\alpha_{jk}p_{jk} \right\}.
\end{eqnarray}
This relay assignment has to be followed with solving $J$ water-filling
problems to meet the constraint of a minimum rate to each user. Note
that it is possible that the simplified optimization problem is
feasible where as the original optimization problem in
Eqns.~\eqref{maxwithminobj}-\eqref{maxwithminrate} is not. It is also
possible that solving the $J$ water-filling problems to compute the
lower bound might be an infeasible optimization problem. In these
cases, the bounds are not meaningful. However, these scenarios occur
very rarely.

\subsection{Max-min rate }
We will now consider a third metric, the minimum rate to each user. The
optimal power allocation ensures that each user receives the same data
rate. The optimization problem maximizes the minimum rate to all the
users subject to two constraints: only a single relay helps each user
and each relay must meet a power constraint. The optimization problem
is,
\begin{eqnarray}
 \max_{\left\{\alpha_{jk}\right\}} \min_k \left\{ \frac{1}{2} \log_2 \left( 1+c_k +
                    \sum_{j=1}^J p_{jk} \alpha_{jk} \right) \right\} \label{maxminobj} \\[10pt]
\mathrm{such\ that } \hspace*{0.3in} \forall k,\, \alpha_{jk}\alpha_{lk} = 0, \hspace*{0.3in}
                               j\neq l,j,l \in \left\{1, 2, \dots, J\right\} \label{maxminconst} \\
                               [10pt]
\sum_{k=1}^K \alpha_{jk} = 1 \hspace*{0.3in} \forall \hspace*{0.1in} j, \label{maxminpow} \\[10pt]
\alpha_{jk} \geq 0 \hspace*{0.3in} \forall \hspace*{0.1in} k,j. \label{maxminpositivepower}
\end{eqnarray}

As before, ther than Eqn.~\eqref{maxminconst}, the optimization problem
in Eqns.~\eqref{maxminobj}-\eqref{maxminpositivepower} is concave: the
objective function is concave and the remaining constraints are
affine~\cite{Boydbook}. As before, we ignore the constraints given in
Eqn.~\eqref{maxminconst} and maximize the objective function subject to
the power constraints alone:
\begin{eqnarray}
\max_{\left\{\alpha_{jk}\right\}} R & = &  \max_{\left\{\alpha_{jk}\right\}}
                    \min_{k} \left\{ \frac{1}{2} \log_2 \left( 1+c_k +
                           \sum_{j=1}^J p_{jk} \alpha_{jk} \right) \right\}, \label{maxminobj1} \\
\mathrm{such\ that} & & \sum_{k=1}^K \alpha_{jk} = 1 \hspace*{0.3in}
                                                \forall \hspace*{0.1in} j, \label{maxminpow1} \\
& & \alpha_{jk} \geq 0 \hspace*{0.3in} \forall \hspace*{0.1in} k,j. \label{maxminpositivepower1}
\end{eqnarray}
Note that the objective function given by Eqn.~\eqref{maxminobj1} is
not differentiable. To analyze this problem, we formulate an equivalent
optimization problem with differentiable objective and constraint
functions.

The logarithm function is a monotonically increasing
function of its argument, and hence, maximizing the minimum of
logarithm functions is same as maximizing the minimum of the arguments
of the logarithm function. Also, for any positive real numbers
$x_1,x_2,\ldots,x_n$,
$$ \min_i \left\{ x_i \right\} = \left( \max_i \left\{ \frac{1}{x_i} \right\} \right)^{-1}
\hspace*{0.1in} \mathrm{and} \hspace*{0.2in} \max_i \left\{ x_i \right\} =
\lim_{l \rightarrow \infty} \left( \sum_{i=1}^{n} x_i^l \right)^{1/l}$$

Using these relations, the objective function can be reformulated as
\begin{equation}
\min_{\left\{\alpha_{jk}\right\}} \left(\frac{1}{R}\right) =
\min_{\left\{\alpha_{jk}\right\}} \left[ \lim_{l \rightarrow \infty}
        \left( \sum_{k=1}^{K} \frac{1}{ \left( R_{k}^J \right)^l} \right)^{1/l} \right],
                                                                            \label{eq:linfobj}
\end{equation}
where $R_{k}^J=1+c_k + \sum_{j=1}^J p_{jk} \alpha_{jk}$.

Consider the optimization problem for some finite $l$,
\begin{eqnarray}
\min_{\left\{\alpha_{jk}\right\}} (\frac{1}{R}) & = & \min_{\left\{\alpha_{jk}\right\}}
           \left( \sum_{k=1}^{K} \frac{1}{ \left( R_{k}^J \right)^l} \right)^{1/l}, \label{egobj}\\
\mathrm{such\ that} & & \sum_{k=1}^K \alpha_{jk} - 1 = 0 \hspace*{0.3in} \forall
                                                            \hspace*{0.1in} j,\label{egpow}\\
& & - \alpha_{jk} \leq 0 \hspace*{0.3in} \forall \hspace*{0.1in} k,j. \label{egpositivepower}
\end{eqnarray}

Again we show that the solution to the relaxed problem leads to
selection in most cases. Note that while these conditions are the same
as derived earlier for the max-sum rate, the approach to this getting
here is very different. The optimization problem given by
Eqns.~\eqref{egobj}-\eqref{egpositivepower} is also a convex
optimization problem\cite{Boydbook}. To show that when the power is
optimally allocated, most users receive power only from one of the
relays, let us characterize the set of solutions to the optimization
problem. For the sake of clarity, we again start by looking at the case
with $J=2$ relays. In such a case, the Lagrangian of the minimization
problem is given by
\begin{eqnarray}
\mathcal{L}(\left\{\alpha_{1k},\alpha_{2k}\right\},\{\lambda^{1}_k\},\{\lambda^{2}_k\},\nu_1,\nu_2)
    &=& \left( \sum_{k=1}^{K} \frac{1}{ \left( R_{k}^2\right)^l} \right)^{1/l} +
    \sum_{k=1}^{K}\lambda^{1}_k (-\alpha_{1k})
  +\sum_{k=1}^{K}\lambda^{2}_k (-\alpha_{2k}) \nonumber \\
 & & + \nu_1 \left(\sum_{k=1}^{K} \alpha_{1k} -1\right)
 + \nu_2 \left(\sum_{k=1}^{K} \alpha_{2k} - 1\right),
\end{eqnarray}
where $R_{k}^2=(1+c_k + p_{1k} \alpha_{1k} + p_{2k} \alpha_{2k}) $;
$\lambda^{1}_k$ and $\lambda^{2}_k$, $k = 1,2,\ldots, K$ are the
Lagrange multipliers associated with the constraint of positive power
given by Eqn.~\eqref{egpositivepower}; and $\nu_1$ and $\nu_2$ are the
Lagrange multipliers associated with the total power constraints  given
by Eqn.~\eqref{egpow}. The KKT conditions, which must be satisfied, are
\begin{eqnarray}\label{egkkt}
\hspace*{-2.8in} \sum_{k=1}^{K} \alpha_{1k}=1, \qquad \sum_{k=1}^{K} \alpha_{2k}=1
\qquad \qquad \\[10pt]
\hspace*{-2.8in} -\alpha_{jk}\leq 0 \qquad \forall j, k \qquad \qquad \qquad \qquad  \\[10pt]
-f_{R}p_{1k} - \lambda^{1}_k + \nu_1 = 0, \quad \lambda^{1}_k \alpha_{1k} = 0,
\quad \lambda^{1}_k \geq 0, \quad \forall k,\\
-f_{R}p_{2k} - \lambda^{2}_k + \nu_2 = 0, \quad \lambda^{2}_k \alpha_{2k} = 0,
\quad \lambda^{2}_k \geq 0, \quad \forall k,
\end{eqnarray}
where $$f_{R}=\left( \sum_{k=1}^{K} \frac{1}{ \left( R_{k}^2\right)^l}
\right)^{\frac{1}{l}-1} (R^{2}_k)^{-l-1}.$$

Now suppose for some $i\in \{1,2,\ldots,K\}$, $\alpha_{1i}$ and $\alpha_{2i}$ are both
non-zero, then the conditions $\lambda^{1}_i \alpha_{1i} = 0 $ and $\lambda^{2}_i \alpha_{2i} = 0$
dictate that  $\lambda^{1}_i$ and $\lambda^{2}_i$ are both zero.
>From the KKT conditions it follows that
\begin{equation}\label{egconst1}
\frac{\nu_1}{p_{1i}}=\frac{\nu_2}{p_{2i}}.
\end{equation}
As discussed earlier, no other pair ($\alpha_{1k},\alpha_{2k}$) can
have two non-zero entries. Note that this property of the solution is
true for all $l$. Therefore, the solution to the optimization problem
with the objective function given by Eqn.~\eqref{eq:linfobj}, and
constraints given by
Eqns.~\eqref{maxminpow1}-\eqref{maxminpositivepower1}, also has this
property. The objective function given by Eqn.~\eqref{eq:linfobj} is a
reformulation of Eqn.~\eqref{maxminobj1}, and, in turn, the solution to
the optimization problem given by equations
Eqns.~\eqref{maxminobj1}-\eqref{maxminpositivepower1} also has the
aforementioned property for the case of two relays. Therefore, for the
case of two relays, the solution obtained by ignoring
Eqn.~\eqref{maxminconst} comes quite close to the solution to the
original optimization problem given by
Eqns.~\eqref{maxminobj}-\eqref{maxminpositivepower}.

To generalize this result, the solution to the max-min optimization
problem given by Eqns.~\eqref{maxminobj1}-\eqref{maxminpositivepower1},
for the case of $J$ relays has at most $J-1$ of the $J$-tuples
($\alpha_{1k}, \alpha_{2k}, \ldots, \alpha_{Jk}$) with more than one
non-zero entry.

The construction of a heuristic to the solution of the optimization
problem given by Eqns.~\eqref{maxminobj}-\eqref{maxminpositivepower},
is, as before: assign each user receiving power from multiple relays to
that relay from which it receives  maximum power. If there are multiple
relays which allot the same maximum power, assign the user to any one
of them arbitrarily. Once this relay assignment is done, if required,
$J$ max-min power allocation algorithms are solved for the power
distribution at each of the relays.

\subsection{Independent codebooks at the relays}
In the previous section, relay selection and power allocation was done
for the case when the transmitter and the relays use the same codebooks
to encode the messages. The results can also be extended to the case
when independent codebooks are employed at the source and the relays.
Using independent codebooks results in higher rates\cite{Laneman_dstc},
however, decoding of the source and relay messages is significantly
more complex compared to the case of repetition
coding\cite{Implementingcoop}. When the source and the relays employ
independent Gaussian codebooks, the optimization problem to maximize
the sum rate to all users, similar to the ones given in equations
\eqref{mainobj}-\eqref{pow}, is given by:
\begin{eqnarray}
\max_{\left\{\alpha_{jk}\right\}} R & = &  \max_{\left\{\alpha_{jk}\right\}}
                  \sum_{k=1}^{K} \Bigg{\{} \frac{1}{2} \log_2 \left( 1+c_k \right) +
      \frac{1}{2} \log_2 \left( 1+\sum_{j=1}^J p_{jk} \alpha_{jk} \right) \Bigg{\}},  \\[10pt]
\mathrm{such\ that} & \forall k, & \alpha_{jk}\alpha_{lk} = 0, \hspace*{0.3in}
                 j\neq l,j,l \in \left\{1, 2, \dots, J\right\} \label{indep_const}  \\[10pt]
& & \sum_{k=1}^K \alpha_{jk} = 1 \hspace*{0.3in} \forall \hspace*{0.1in} j, \\[10pt]
& & \alpha_{jk} \geq 0.
\end{eqnarray}

It is not hard to show that other than the constraint given in
Eqn.~\eqref{indep_const}, the optimization problem is a concave
maximization problem, and like before, solving it gives an upper bound
to the sum rate. The heuristic which also serves as a lower bound can
also be constructed from it. A max-min optimization problem can also be
formulated in a similar manner.

\section{Numerical Results and Discussion} \label{res}

In this section we verify the validity of the assumption in
Eqn.~\eqref{assump} and present the results of simulations to
illustrate the tightness of the bounds developed in the previous
section. We compare the performance of three cases: the baseline
scenario uses a single-input single-output system (SISO) with a single
antenna at the BS and user and relaying is not used. The alternative is
a system with a single antenna at the BS and $J$ relays with a single
antenna each. The last system considered is a multiple-input
single-output (MISO) system with $J+1$ antennas at the BS and a single
antenna at each user. In comparing these cases, all other system
parameters, e.g., number of users, total power and bandwidth, remain
constant.

\subsection{Channel Model}
The simulations are implemented using the COST-231 channel model as
described in \cite{80216jrelay}. The model assumes both the BS and
relays are at some height off the ground and treats the BS-relay
channel as Rician. The BS-destination and relay-destination channels
are modeled as Rayleigh. The path loss in the BS-relay channel is made
up of two components, free space loss and multi-screen loss. In
addition to these two, the BS-user and the relay-user channels have a
rooftop-to-street diffraction loss. For the values of the parameters
that we consider, the COST-231 channel model suggets a distance
attenuation in channel power of 20dB/km for the first 657 meters and
38dB/km for greater distances. The model therefore appears to be
conservative in the sense that one would expect the LoS component in
the Rician fading to attenuate slower than the other non-LoS
components. In the MISO case, the large scale fading in all the
channels between the transmit antennas and a particular user, is the
same. \emph{Each user is assigned an orthogonal channel of bandwidth of
200kHz}, resulting in a noise power of -120dBm. The chosen system
parameters are given in Table~\ref{tab:param}.

\begin{table}
\caption{Parameters used in COST231 model}
\centering
\begin{tabular}{|c|c|c|c|}
\hline
Parameter & Value chosen & Parameter & Value chosen\\
\hline
 BS height			&  50m & Rooftop height & 30m\\
Relay height  &  50m & User height  & 1.5m\\
Frequency		&  1GHz & Road orientation & 90 degrees \\
Building spacing & {50m} & Street width & {12m}\\
Transmit power & 20dBm & Noise power spectral density & -174dBm/Hz\\
\hline
\end{tabular}
\label{tab:param}
\end{table}

\subsection{Decoding at the relays}
To form a tractable problem, we had made the assumption that the relays
always successfully decode the message transmitted by the BS in the
first time slot, and the data rate is the limited by compound
source-relay-destination channel capacity, as in Eqn.~\eqref{assump}.
To verify the assumption, we consider a circular cell, centered at a
BS, of radius one kilometer with $J=4$ relays positioned at ($\pm
200\sqrt{2}m, \pm 200\sqrt{2}m$), i.e., on a ring of radius 400m.
$3\times 10^6$ user locations in the cell are randomly generated. For
each location, independent channels are generated using the
 channel model. As shown in Fig.~\ref{sysmodel}, we divide the cell into annular rings of radius
100 meters. In Table~\ref{loc} we list the percentage of number of
locations where Eqn.~\eqref{assump} is valid. It is evident from the
table that the assumption we make is valid whenever the user is farther
than 300m from the BS. Essentially, for all user locations of interest,
i.e., areas where users have a relatively weak channel to the BS, the
assumption is valid. It is worth emphasizing that these are
conservative numbers.

\begin{table}
\centering
\caption{Percentage of locations where Eqn.~\eqref{assump} is satisfied}
\begin{tabular}{|c|c|c|c|}
\hline
Distance from & $\%$  & Distance from & $\%$ \\
 the BS (m)   &    locations            &  the BS (m)   & locations\\
\hline
0-100 & 93.591   & 500-600 &  99.943\\
100-200 & 99.642 & 600-700 & 99.963\\
200-300 & 99.815 & 700-800 & 99.977\\
300-400 & 99.309 & 800-900 & 99.989\\
400-500 & 99.482 & 900-1000 & 99.992\\
\hline
\end{tabular}
 \label{loc}
\end{table}

\subsection{Tightness of the bounds}

Our next of simulations test the tightness of the upper bound as
developed in this paper and the resulting heuristic which acts as a
lower bound. Note that this heuristic is our final solution to the
joint selection and power allocation problem. The relay assignment and
the power allocation is done based on the instantaneous channel powers.
For this simulation, the channels are generated as independent
realizations of a unit-variance Rayleigh fading random variable. For a
fair comparison, the power allocated to each relay is set to $1/J$,
i.e., all curves use the same total power. The curves presented here
are averages over one thousand random user locations.

\begin{figure}
\centering
 \includegraphics[scale =0.6]{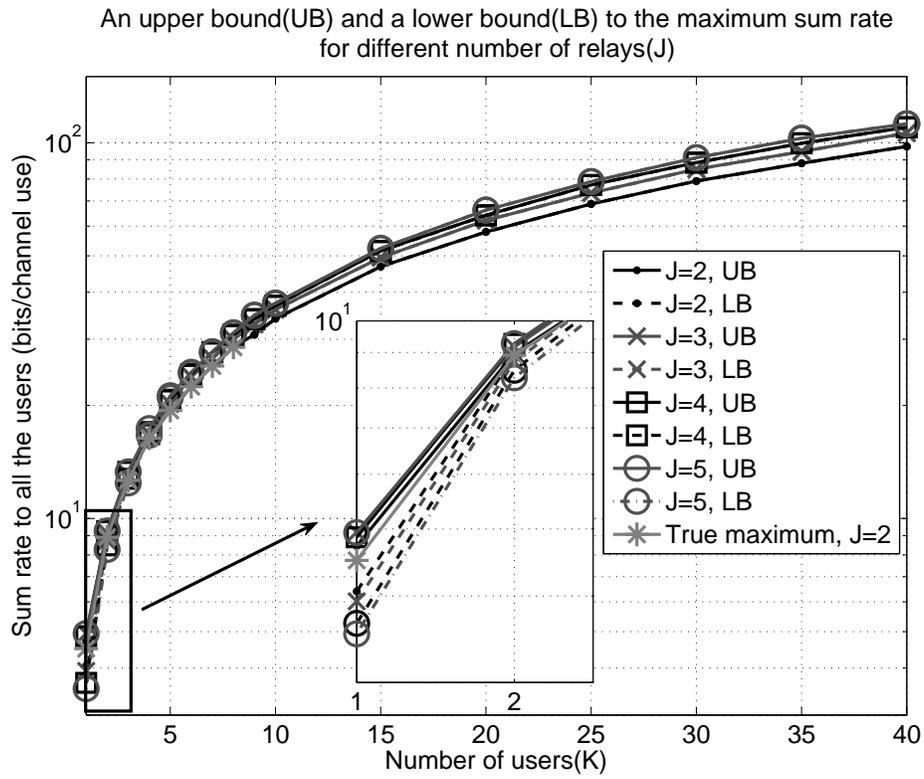}
\caption{The proposed upper bound to the maximum sum rate and the
heuristic (a lower bound) as a function of the number of users. Note that
both the bounds are extremely tight.}
\label{boundmaxsum}
\end{figure}

Figure~\ref{boundmaxsum} plots the upper bound, and the sum rate
achievable by the heuristic (that also acts as a lower bound on the
achievable sum rate) for varying values of $J$ and $K$. The average
total transmit power to noise power ratio is set to 30dB. As is clear
from the figure, the upper and lower bounds are indistinguishable. As
explained in Section~\ref{probform}, this is because it is quite rare
for a user to be allocated power from multiple relays, i.e., selection
is essentially inherent in the approximate solution. The heuristic,
therefore, is an extremely effective solution to the joint selection
and power allocation problem. By an exhaustive search, we also find the
exact maximum sum rate for the case with $J=2$ relays and $K$ between
one and eight.  Note that since each exhaustive search requires
solution of $J^K$ water-filling problems, any larger value of $J$ is
infeasible.

\begin{figure}
\centering
 \includegraphics[scale =0.6]{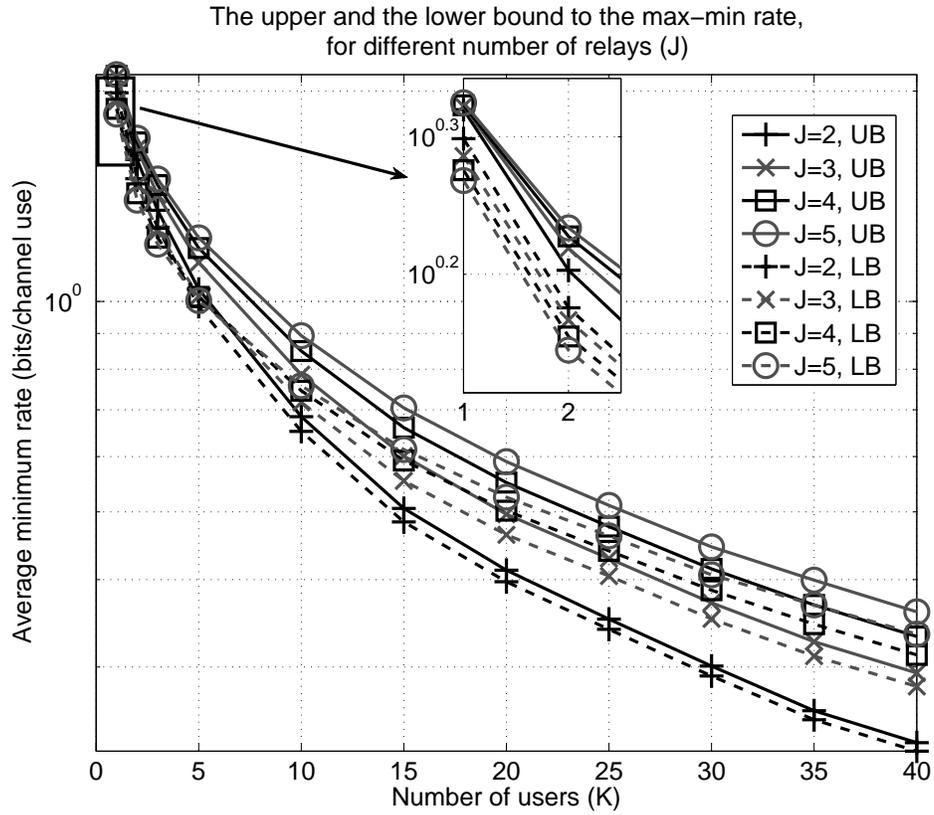}
\caption{The proposed upper and the lowerbound to the max-min rate.}
\label{boundmaxmin}
\end{figure}

Figure~\ref{boundmaxmin} plots the upper and lower bound to the max-min
rate for varying values of $J$ and $K$ averaged over many channel
realizations. In this simulation, the average total transmit power to
noise power ratio is set to 20dB. Again, the bounds are extremely tight
and the heuristic provides an effective solution. The slight difference
is due to the rare case where a user is allocated power by two relays
(see Section~\ref{probform}). In interests of brevity, we do not
provide a similar plot for the max-sum rate with a rate constraint.

\subsection{Results for a cellular network}
In this section, we use the theory developed for solving the max-min
and max sum rate problems, to estimate the performance gains, with
respect to a SISO and MISO system in cellular network setting. We
consider a cell of radius $\mathrm{r}_{\mathrm{cell}}$. Performance
here is measured as the increase in cell-size made possible by
relaying. Since we wish to study the improvement in the rates to the
users with poor channels to the BS, we consider users in the outer
annular ring, of inner radius $\mathrm{r}_{\mathrm{cell}}/2$ and outer
radius $\mathrm{r}_{\mathrm{cell}}$, the area shaded in gray in
Figure~\ref{sysmodel}.  Users are distributed uniformly in the region
with a constant user density of $(30/\pi)$ per square kilometer. We
consider the following three system models for comparison:
\begin{enumerate}
\item A cellular network with a single antenna BS, communicating to
    multiple users with single antenna receivers (multiuser SISO
    system).
\item A cellular network with a BS with five transmit antennas,
    communicating to multiple users with single antenna receivers
    (multiuser MISO system).
\item A cellular network with a BS with a single antenna and
    assisted by four relays positioned on a ring of radius
    $0.4\mathrm{r}_{\mathrm{cell}}$, communicating to multiple
    users with single antenna receivers.
\end{enumerate}

For the simulation, we generate 50 random sets of locations for the
users. We then use the COST-231 model to generate the BS-user and
relay-user channels. For each set of locations, we generate one set of
large-scale fading variables. To average over small-scale fading random
variables, for every set of locations, we generate 500 small-scale
fading random variables. As indicated in Table~\ref{tab:param}, the
total power used in communication is set to 20dBm.

In the first example, powers are allocated to maximize the sum rate to
all the users. For a fair comparison, we use this to compute the data
rate averaged over all users. In the SISO case, the system uses
water-filling to allocate power to the multiple users. In the MISO case,
the BS is assumed to know the channel vector to each user and can both
match to the channel and allocate power via water-filling. Finally, in
the case with relays, selection and power allocation uses the scheme
developed in Section~\ref{probform}.

\begin{figure}
\centering
 \includegraphics[scale =0.6]{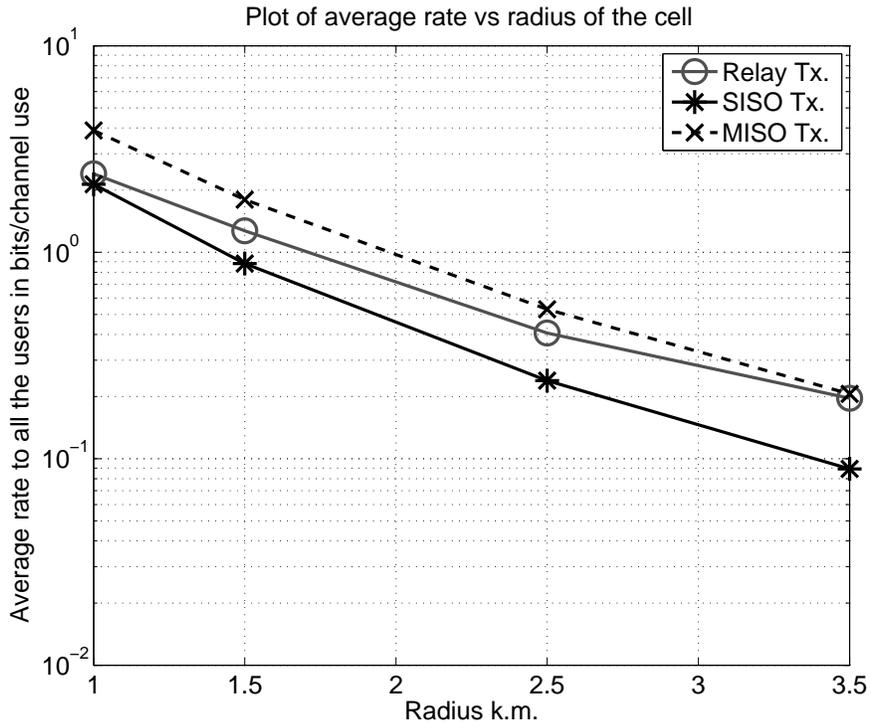}
\caption{Average data rate as a function of the radius of the cell.}
\label{avgrate_maxsum}
\end{figure}

In Figure~\ref{avgrate_maxsum}, we plot the average user rate as a
function of the radius of the cell. We compute the rates as given by
the lower bound, assuming that the power allocation is done only in the
second time slot. In the first time slot, the BS distributes the power
equally among all the users. This is done to ensure that the relays are
able to decode all the transmitted messages. In the second time slot,
each of the relays uses one fourth of the available power to
communicate with the users it assists. This ensures that the total
power used is the same in all the three system models. Interestingly,
Figure~\ref{avgrate_maxsum} shows that a MISO system provides higher
average data rates (and hence the sum rate) compared to the system with
relays. We explain these graphs in the following section.

Next, we repeat the simulations by allocating power using the max-min
algorithm, and then computing the outage rates for each of the system
models. For the SISO and MISO cases, computing this power allocation is
fairly straightforward. The optimal power allocation is the one such
that all the users have the same data rate. When relays are employed,
we use the methodology developed in this paper to solve the max-min
optimization problem.

We plot the outage rates for $10\%$ and $1\%$ outage, as a function of
$\mathrm{r}_{\mathrm{cell}}$ in Figure \ref{outcap_maxmin}. Here we see
a reversal in performances, with  the system with relays providing
higher outage rates compared to the MISO system.  As expected, the
BS-user communication in these systems is more susceptible to channel
fluctuations. This plot is discussed further below.

\subsection{Discussion}
\begin{figure}
\centering
 \includegraphics[scale =0.6]{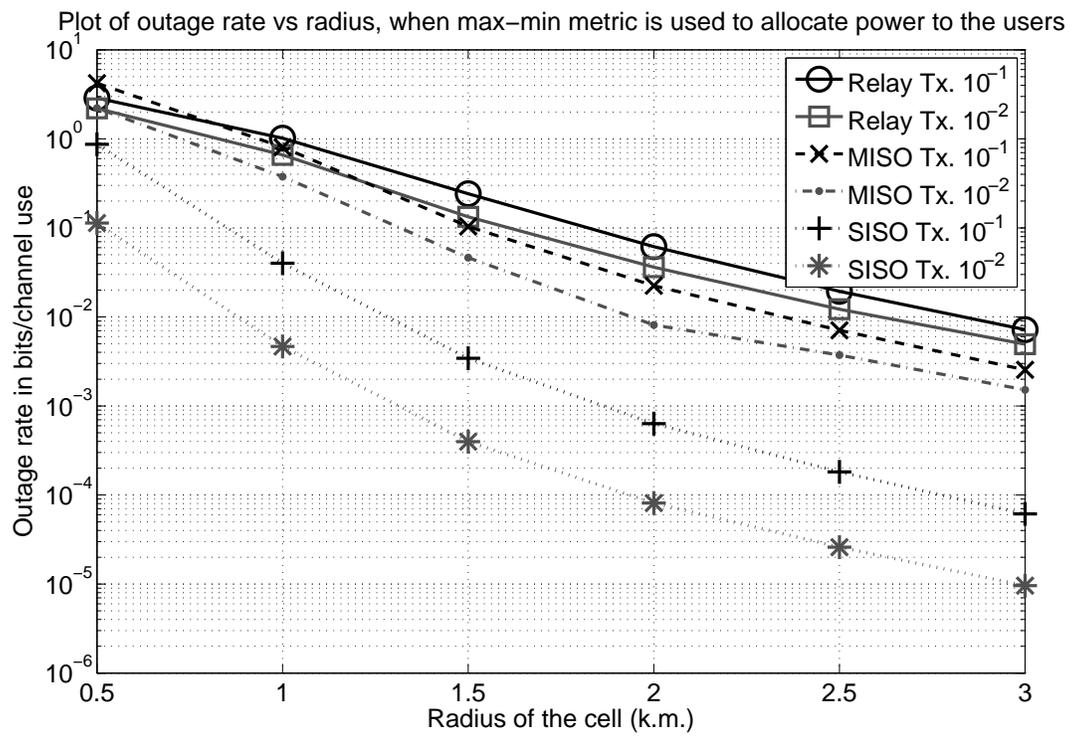}
\caption{Outage rate as a function of the radius of the cell.}
\label{outcap_maxmin}
\end{figure}

A user in a heavily shadowed region has a weak channel to the base
station. Having multiple antennas at the base station does not help
much. Relays aid such users by providing alternate paths to the base
station. This is consistent with the data in Figure~\ref{outcap_maxmin}
where a relay system provides higher outage rates. This is because the
outage rates depend on the data rates to the users with weakest
channels. On the other hand, the max-sum rate algorithm, allocates more
power to the users with the strongest channels. The MISO system
provides higher data rates compared to a relay system. As shown in
Fig.~\ref{avgrate_maxsum}, the loss in half the bandwidth incurred in
switching from direct transmission to co-operative transmission
outweighs the benefits brought by the additional diversity.

In a network setting where a user has the same average channel to all
the $J$ relays, selection cooperation achieves order $J+1$
diversity\cite{Ela_jnl}. However, because of the geometry of a cellular
network and because of the rapid deterioration of the channel powers
with distance, most users have good channels to only a small set of
relays. The \emph{effective} diversity order is,
therefore, limited.

Figures~\ref{avgrate_maxsum} and~\ref{outcap_maxmin} also lead to a
cautionary result. These results indicate that, compared to a system
with SISO communication, deploying relays does offer substantial
improvements. The area serviced effectively by a single BS, helped by
relays, can significantly expand. However, these improvements need to
be commensurate with the infrastructure costs involved in the
deployment of these relays including both the antenna system cost and
`non-technical' costs such as the required real-estate. If the cost of
a relay were on the same order of magnitude as a base station, the
improvements in the cell radius, as shown by the simulations do not
justify the additional cost. Also, depending on the performance metric,
a MISO system may perform better or almost as good as the relay system,
but with significantly lower costs.

Clearly, a complete financial cost/benefit analysis is beyond the scope
of this paper. Furthermore, the examples presented here are limited and
do  not explore every potential parameter. However, do note that the
our results are optimistic in assuming the relays can always decode and
that the transmitters have all the information they need to make
optimal decisions. Our goal here is to indicate that significant gains
are possible, but are context and scenario dependent. These results
also indicate the need for exploring alternate ways of exploiting
cooperative diversity. We also need to explore alternate hybrid schemes
wherein the relays help only those users who need it.

\section{Conclusion}\label{sec:conc}

This paper deals with the use of cooperation in a cellular network
wherein a base station is assisted by a few dedicated relays. Previous
work largely for mesh networks has shown the importance of
\emph{selection}, i.e., each user using only one relay, since this
minimizes the overhead due to orthogonal channels. However, in a
scenario with multiple data flows, selection has been either brute
force or ad-hoc. Previous work has also largely ignored the problem of
power allocation once the selection is achieved. In this paper we
developed an optimization framework to solve the problem of joint
selection and power allocation.

The optimization problem uses the achievable sum rate and max-min
user-rate as two figures of merit. Given that the selection problem has
exponential complexity, in this paper we formulate alternative convex
optimization problems whose solution provides upper bounds on the two
metrics. However, for practical values of number of users, the bound is
indistinguishable from the true solution. Since this solution can
violate the selection condition, a related heuristic is derived that
assigns users to the relay which allocates it the maximum power. The
resulting lower bound is also extremely tight and we have an efficient
solution to the problem at hand. The numerical examples, using
realistic channel models, illustrate the benefits achievable due to
relaying.

\clearpage

\bibliographystyle{IEEEtran}
\bibliography{ref}

\clearpage

\end{document}